# Attosecond dispersive soft X-ray absorption fine structure spectroscopy in graphite

Bárbara Buades[1], Dooshaye Moonshiram[2], Themistoklis P. H. Sidiropoulos[1], Iker León[1], Peter Schmidt[1], Irina Pi[1], Nicola Di Palo[1], Seth L. Cousin[1], Antonio Picón[1,3], Frank Koppens[1,4], Jens Biegert[1,4,*]

[1]ICFO-Institut de Ciencies Fotoniques, The Barcelona Institute of Science and Technology, 08860 Castelldefels (Barcelona), Spain

[2]Institute of Chemical Research of Catalonia (ICIQ-BIST), Avinguda Països Catalans 16, 43007 Tarragona, Spain

[3]Grupo de Investigación en Aplicaciones del Láser y Fotónica, Departamento de Física Aplicada, University of Salamanca, E-37008 Salamanca, Spain

[4]ICREA, Pg. Lluis Companys 23, 08010 Barcelona, Spain

*e-mail: jens.biegert@icfo.eu

**Phase transitions of solids and structural transformations of molecules are canonical examples of important photo-induced processes, whose underlying mechanisms largely elude our comprehension due to our inability to correlate electronic excitation with atomic position in real time. Here, we present a decisive step towards such new methodology based on water-window-covering (284 eV to 543 eV) attosecond soft X-ray pulses that can simultaneously access electronic and lattice parameters via dispersive X-Ray absorption fine-structure (XAFS) spectroscopy. We validate attoXAFS with an identification of the $\sigma^*$ and $\pi^*$ orbital contributions to the density of states in graphite simultaneously with its lattice's four characteristic bonding distances. This work demonstrates the concept of attoXAFS as a powerful real-time investigative tool which is equally applicable to gas-, liquid- and condensed phase.**

X-ray absorption fine-structure (XAFS) spectroscopy is a powerful element-specific technique, providing electronic as well as structural, and chemical information with atomic resolution[1–3]. In XAFS, electronic information is extracted from the near-edge XAFS (XANES or NEXAFS), which arises from transitions from inner-shell orbitals to unoccupied electronic states near the Fermi energy level. Its measurement requires high spectral resolution to resolve its features that occur within only a few eV. Structural information is mainly obtained from the extended XAFS (EXAFS), which extends over several hundred eV above the absorption edge, and it arises from the scattering of photo-emitted core-electrons on neighboring atoms. While XANES and EXAFS are both well-established methods, and quick-scan or dispersive setups permit a relatively rapid acquisition of EXAFS[4,5], crucially lacking so far was the capacity to connect electronic with structural information in real-time, i.e. faster than the typical tens of femtosecond dephasing time of electronic wavepackets[6]. The capability to probe the connection between a material's electronic excitation and lattice reordering provides a new powerful tool to gain insight into the real-time evolution non-equilibrium dynamics such as e.g. structural and electronic phase transitions.

Here, we achieve this key requirement through the simultaneous measurement of XANES and EXAFS with an attosecond soft X-ray continuum (attoXAFS). Until now, the limiting factor for an implementation of attoXAFS was the broadband spectral coverage and short temporal duration of X-ray pulses. Thus, with the advent of high-harmonic generation (HHG) light sources[7,8], novel tools for XAFS have emerged which can provide the desired attosecond temporal resolution with sufficient shot-to-shot reproducibility on a tabletop. However, it is only recently that isolated attosecond, i.e. ultrabroadband, pulses in the soft X-ray (SXR) water-window (284 eV to 534 eV) with sufficient flux have come online[9–11], thus providing the desired ultrafast temporal resolution in combination with sufficient spectral coverage in a photon energy range that offers element specificity.

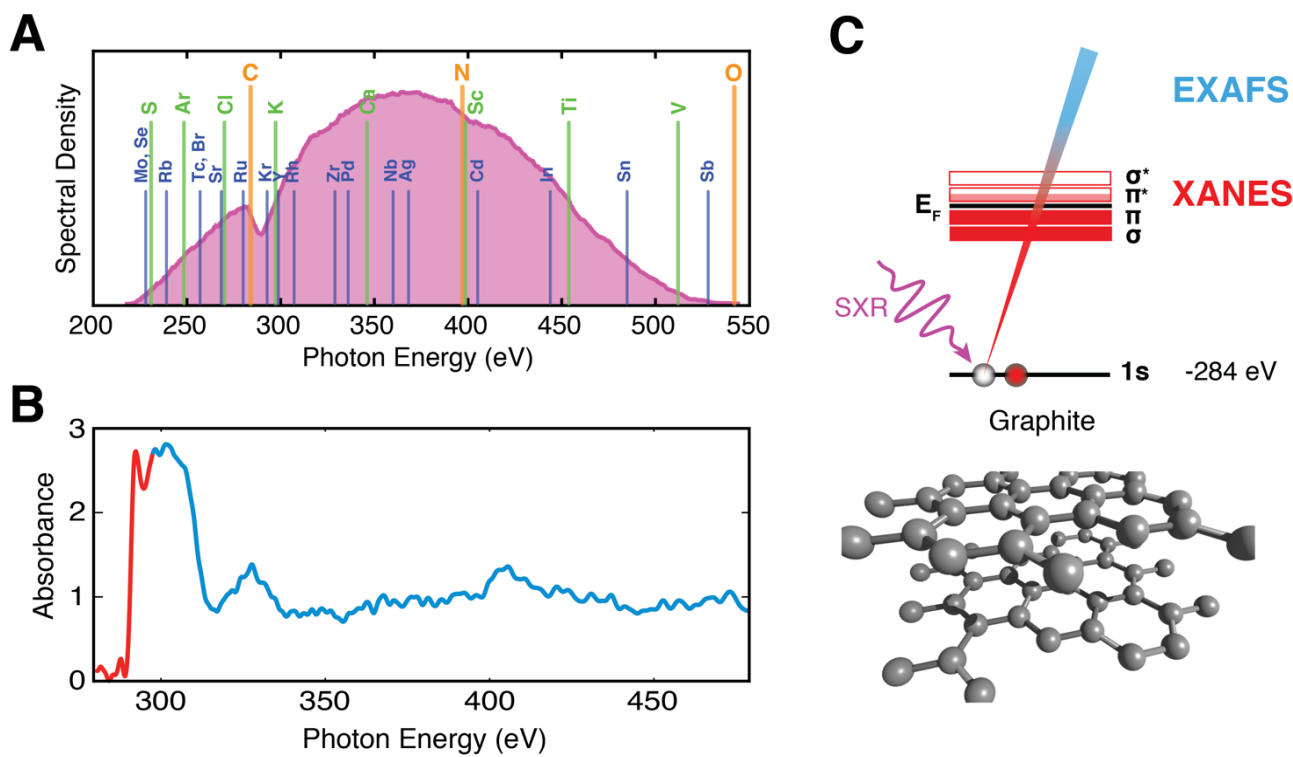

**Figure 1. AttoXAFS - Simultaneous XANES and EXAFS with an isolated attosecond soft X-ray pulse.**

(A) Shown is the spectrum of the isolated attosecond pulse spanning the entire soft X-ray water window (284 – 543 eV) at once. Overlaid as vertical lines are the positions of accessible K (orange), L (green) and M-shell (blue) absorption edges. (B) shows the attoXAFS measurement in graphite (structure shown bottom right) in which the XANES is highlighted in red and the EXAFS in blue. Note that no hard boundary exists between XANES and EXAFS and that the color separation is solely chosen to indicate the predominant contribution to the measurement. (C) indicates that spectral components with photon energy above 284 eV access 1s carbon K-shell core electrons. Transitions to bound states close to the Fermi edge (red) are known as XANES and transitions into the continuum (blue) give rise to the EXAFS. XANES provides orbital sensitivity depending on the angle of the impinging X-ray pulse field vector relative to the basal plane normal of the sample while EXAFS predominately provides local structural information as it is related the scattered electron wave from the absorber atom on the neighboring atoms.

Here, we demonstrate the concept of attoXAFS in graphite with an isolated attosecond water-window SXR pulse that enables the simultaneous acquisition of carbon K-edge XANES and EXAFS, thus providing combined electronic and structural information. The spectrum supporting the sub-300-as-duration pulse[10] is shown in Fig. 1A it covers a region 50 eV below the carbon K-edge (284 eV) up to the oxygen K-edge (543 eV), thus including the K-edge of nitrogen (410 eV) as well as many L and M edges[10] that are relevant to biology and material science. The linearly polarized attosecond pulse was generated via high-harmonic generation and provided adequate photon flux of $7.3 \cdot 10^7$ photons/s for the 320-eV bandwidth; details regarding the attosecond source can be found in Refs.[9–12]. The attosecond SXR pulse was focused with an elliptical mirror (Zeiss) onto a 95-nm-thin graphite sample which was mounted on a 50-nm $Si_3N_4$ substrate. The transmitted signal was analysed by a homebuilt SXR spectrograph, which consisted of a flat-field imaging grating (Hitachi, 2400 lines/mm) and a Peltier-cooled, back-illuminated charge-coupled device (CCD) camera (PIXIS-XO-2048B, Princeton Instruments). To maximize photon flux on the CCD, the sample plane coincided with the entrance slit plane of the spectrograph, thus, without any entrance slit, the SXR focal spot determined the spectrograph's resolution to 1/1000, i.e. 0.3 eV at 300 eV. Figure 1B shows the measured absorbance from dispersive attoXAFS, which is determined according to $-\log[S/S_0]$, where $S$ is the spectral intensity through the sample, and $S_0$ the reference signal measured through only the 50-nm $Si_3N_4$ membrane. Clearly visible in the attoXAFS are sharp absorption features in the XANES region of the spectrum (Fig. 1B,C red) and oscillations in the EXAFS region (Fig. 1B,C, blue).

Graphite consists of equally-oriented graphene layers, which interact via Van-der-Waals forces. In each graphene layer, the carbon atoms form strong covalent bonds and are orientated in a hexagonal lattice with four carbon atoms per unit cell. Each carbon atom has six electrons of which two are located at the 1$s$ core shell and the other four electrons occupy the valence shells ($2s^2$, $2p^2$). Three valence shell electrons form covalent bonds in graphite through a $sp^2$ hybridization. The fourth valence electron, has p symmetry with an orientation perpendicular to the graphene plane ($p_z$ hereafter) and binds via Van-der-Waals forces to the neighboring layers. The $sp^2$ orbitals form σ states in the sheet plane, and $p_z$ forms a π state, which is the highest occupied state[13,14]. The antibonding states σ* and π*, are unoccupied in graphite and form the conduction band. The different symmetries of the two antibonding states, σ* and π*, become apparent when comparing the density of states (DOS) of the two different bands.

XANES at the carbon K-shell edge results from dipole-allowed transitions from the 1s state to the lowest unoccupied states, σ* and π*. Figure 2A shows the XANES part of the normalized attoXAFS spectrum taken at different incident angles of the linearly-polarized attosecond SXR pulses with respect to the graphite basal plane normal. What we found, independent of the incident angle, is a clear rising edge at (292.6 ± 0.3) eV, which is followed by undulations of the absorption spectrum. Figures 2B and 2C show the orbitals calculated with the density functional theory (DFT) code BAND[15,16]; see methods for further information. From symmetry considerations, it is apparent that the edge arises due to the $1s \rightarrow \sigma^*$ transition, indicated in Fig. 2C. Varying the incidence angle from 0° to 20°, a second peak appears at (285.5 ± 0.3) eV with an amplitude that increases as the incident angle is further increased to 40°. The clear field polarization dependence of the peak at 285.5 eV allows us to identify this as $1s \rightarrow \pi^*$ transition since the π* is formed by $p_z$ that is orientated perpendicular to the sample plane. At normal incidence, graphite's plane is perpendicular to the beam propagation direction, thus the attosecond pulse's linearly-polarized electric field probed only in-plane states, σ*, consisting of $sp^2$. By tilting the sample, hence changing the angle of incidence from 0 to 40 degrees, an electric field component along $p_z$ is induced, which excites the $1s \rightarrow \pi^*$ transition. We find that these features, measured with the attosecond source, are in excellent agreement with predicted values as well as with electronic transitions[17–19] measured at synchrotron light sources. We note that the identified features are different from optical absorption measurements in which the $\pi \rightarrow \pi^*$ transition is only excited for in-plane polarization[13], thus the difference with an X-ray measurement arises from the different initial states leading to different symmetry-allowed final states. The attoXAFS measurement accuracy is also sufficient to identify the weak feature, between the π* and σ* resonances at about 289 eV, as interlayer state arising from small residual material impurities[19].

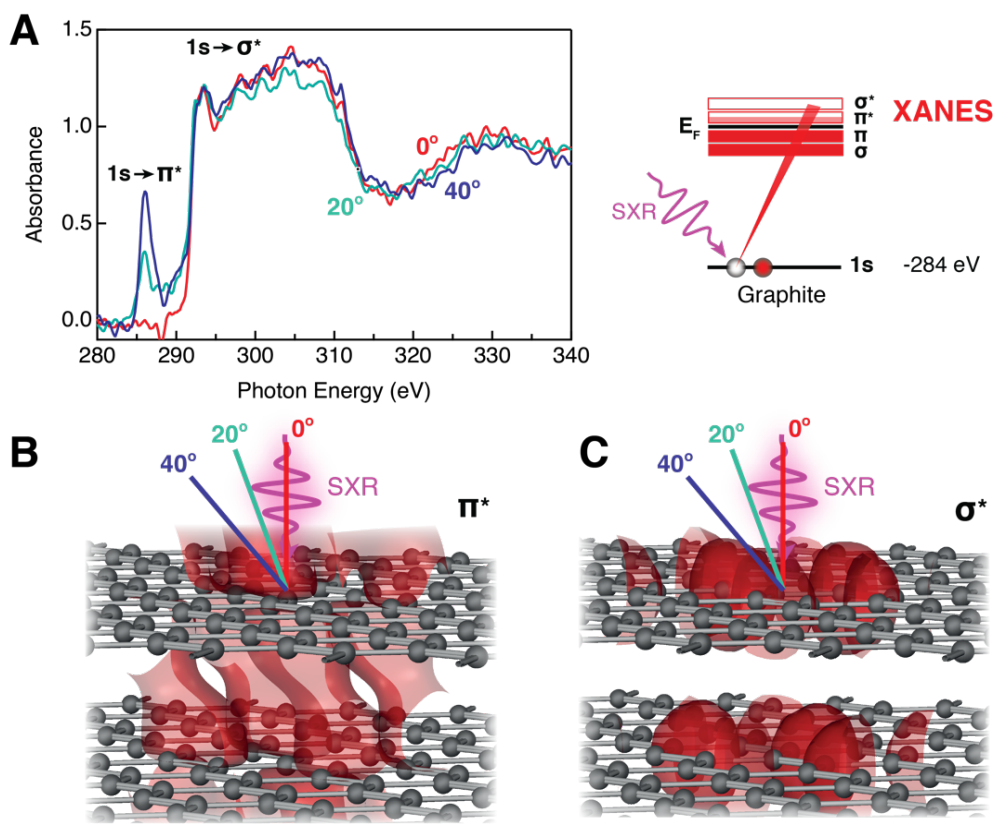

**Figure 2. AttoXAFS identifies different electronic orbitals**. (A) shows the XANES region of attoXAFS for three measurements in which the angle of incidence of the impinging attosecond SXR pulse is varied with respect to the basal plane normal of the sample. Normal incidence (shown also in (B) and (C)) probes states in the basal plane whilst any deviation from normal incidence also probes components out of the basal plane. (B) and (C) show electronic orbitals from DFT calculations using BAND[15,16]. The DOS correspond to the lowest unoccupied states π* (B) and σ* (C) in the conduction band. Two main features can be identified from (A) at 285.5 eV and 292.5 eV, corresponding to $1s \rightarrow \pi^*$ and $1s \rightarrow \sigma^*$ transitions, respectively.

Next, we turn to analyzing the EXAFS region to retrieve information about graphite's lattice conformation, shown in Fig. 3A. We recall that EXAFS arises from the interference of excited photoelectrons with backscattered photoelectrons off neighboring lattice atoms leading to constructive and destructive interference of the absorbing atom's electronic density and thus to oscillations in the measured extended absorption spectrum. The blue part of the measured absorption curve, shown in Figure 1B, displays the EXAFS of graphite. We deduce the bond length from these oscillations with aid of the Athena and Artemis software packages[20] and perform multiple scattering path simulations with FEFF[21]; details are found in the methods section. In short, after background correction and conversion into wavenumber space, shown in Fig. 3B, curve fitting is performed based on the EXAFS equation with scattering phases and amplitudes from FEFF. Figure 3C shows the transformed amplitudes in which we identify three prominent peaks corresponding to the scattering contributions from the $1^{st}$, $2^{nd}$, $3^{rd}$ and $4^{th}$ neighboring carbon atoms. Note that while only the amplitudes of the Fourier transform to R space are shown in Fig. 3C, our fit also takes the phase shifts into account, thereby resulting in a fitting accuracy of better than 2%. From our fit we identify the first apparent peak at 1.35 Å which, in accordance with Refs.[22,23], arises due to the scattering with the nearest neighbor at around 1.42 Å - 1.44 Å. The second peak represents the summed contribution of the second and third neighbor atoms at 2.47 Å and 2.87 Å with coordination numbers 6 and 3, respectively; see supplementary information. Lastly, the third peak represents the scattering from the forth nearest carbon atom at around 3.77 Å.

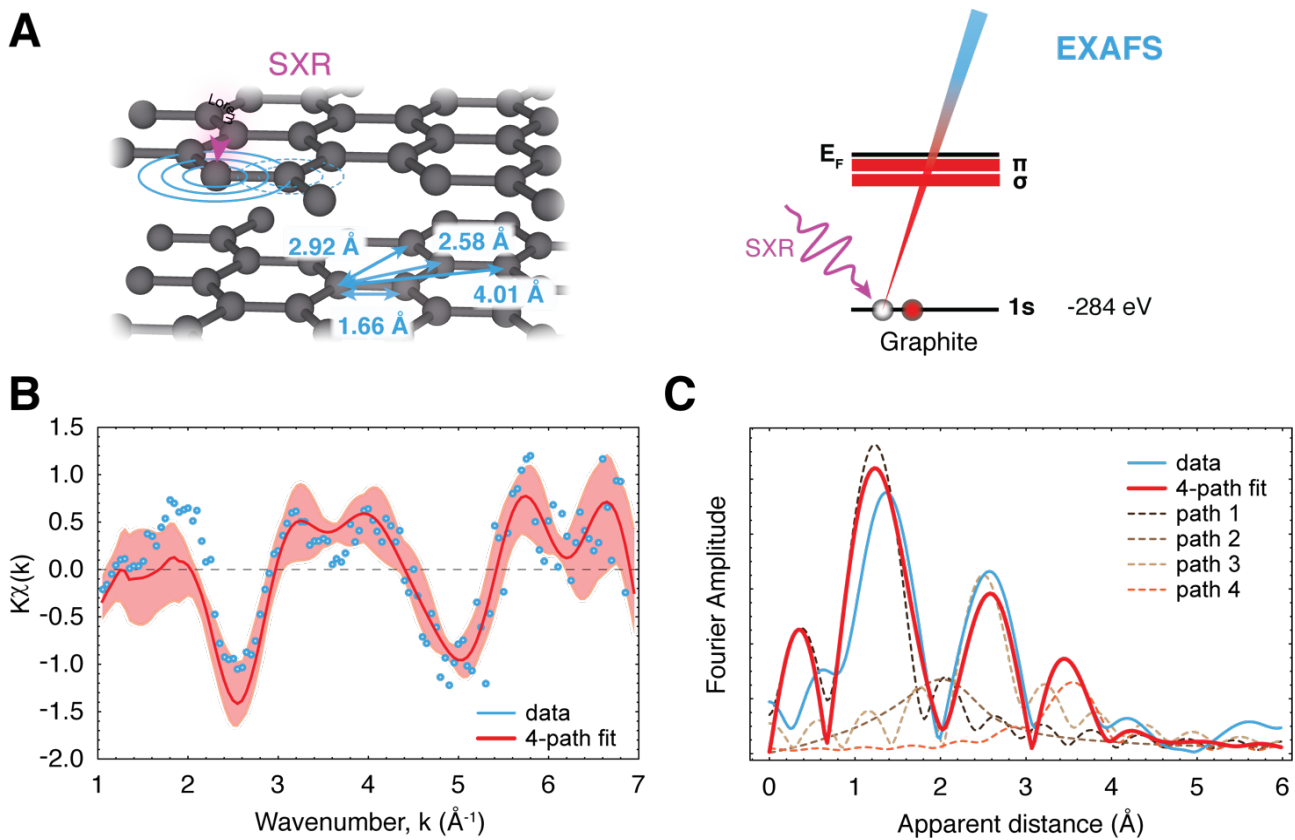

**Figure 3. Bond distances from attoXAFS.** (A) shows two layers of graphite together with the identified bond distances. The experimental EXAFS is shown by the blue circles in (B) together with the back transformed fit (solid red line) and the fit's uncertainty (red shaded area). (C) shows the Fourier amplitude of the experimental data together with the individual contributions from the first 4 scattering paths (dashed lines). The summed scattering contribution is shown in by the solid red line and, considering the scattering phases, results in a 2% fit accuracy. Note that the experimental spectra correspond to k values of 1.18 to 6.90 Å$^{-1}$.

Finally, the analysis of the overall spectrum yields bond distances of (1.66 ± 0.03) Å, (2.58 ± 0.12) Å, (2.92 ± 0.03) Å and (4.01 ± 0.10) Å. These bond distances from attoXAFS are identified in Fig. 3A and in excellent agreement with DFT calculations and carbon K-edge EXAFS measurements conducted at synchrotron sources[22].

In conclusion, we demonstrate with attoXAFS a powerful new tabletop methodology to simultaneously probe electronic states and atomic positions in condensed phase. We validate attoXAFS through a simultaneous identification of the $\sigma^*$ and $\pi^*$ orbitals in graphite in synchronicity with the material's four characteristic bonding distances. Further, the change of angle of incidence of the attosecond pulse permits distinguishing the various orbital contributions to the density of state of the material. The *simultaneous interrogation* of electronic states and lattice conformation permits the *in-situ* identification of electron-phonon coupling and to image how multi-body correlations arise during a material's phase transition or the formation of the superconducting state. This work thus provides a powerful real-time investigative tool to access the characteristic time-scale of electronic motion to resolve charge migration, electron-electron correlation, electron-nuclear scattering and structural transitions. Finally, the access to core level states in synchronicity with valence and conduction band states enables the investigation of core, valence and lattice correlations on a previously unprecedented attosecond time scale.

**Acknowledgements**

We acknowledge financial support from the Spanish Ministry of Economy and Competitiveness (MINECO), through the "Severo Ochoa" Programme for Centers of Excellence in R&D (SEV-2015-0522), the Catalan Institució Catalana de Recerca I Estudis Avançats, Agencia de Gestió d'Ajuts Universitaris i de Recerca (AGAUR), the Fundació Cellex Barcelona, and LASERLAB-EUROPE (EU-H2020 654148). B.B. was partially supported by the PhD fellowship program Severo Ochoa (2013) for doctoral training, awarded by the Spanish MINECO. I.L. was partially supported by MINECO for a Juan de la Cierva postdoctoral fellowship. P.S. was partially supported by a scholarship from the La Caixa Banking Foundation, I.P was partially supported by the Fundació Catalunya - La Pedrera · Ignacio Cirac Program Chair and A.P. was supported by the Marie Sklodowska-Curie Grant

Agreement No. 702565. We thank I. Pi for help with Athena and Artemis and Dr. E. Pellegrin, Dr. J. Herrero-Martin and Prof. Jacinto Sa for discussions related to XAFS.